\documentclass{elsart5p}
\usepackage[english]{babel}
\usepackage{graphicx}
\usepackage{amssymb}

\def \mb{\begin{displaymath}} 
\def \me{\end{displaymath}} 
\def \eb{\begin{equation}} 
\def \ee{\end{equation}}   
\def\expect#1{\mathinner{\langle{#1}\rangle}}

{\catcode`\|=\active 
  \gdef\expect#1{\left<\mathcode`\|"8000\let|\bravert {#1}\right>}}
\def\bravert{\egroup\,\vrule\,\bgroup}

\begin{document}
\begin{frontmatter}



\title{Vibrational effects on low-temperature properties of
molecular conductors}
%

\author[AA]{Jernej Mravlje \corauthref{Mravlje}},
\author[BB,AA]{Anton Ram\v{s}ak},
\author[AA]{Rok \v{Z}itko}

\address[AA]{Jo\v{z}ef Stefan Institute, Jamova 39, Si-1000, Ljubljana, Slovenija}
\address[BB]{Faculty of Mathematics and Physics, University of
  Ljubljana, Jadranska 19, Si-1000, Ljubljana,
Slovenija}

\corauth[Mravlje]{Corresponding author. E-mail address: jernej.mravlje@ijs.si,
Fax:~+38661~477-3724. }

\begin{abstract}
We calculate characteristic correlation functions for the Anderson
model with additional phonon-assisted coupling to the odd conduction
channel.  This model describes, for example, the behavior of a
molecule embedded between two electrodes in linear transport
experiments where the position of the molecule with respect to the
leads affects the tunneling amplitudes. We use variational projection-operator method and
numerical renormalization group (NRG) method. The spin is Kondo screened either by even or odd
conduction channel depending on the gate voltage and electron-phonon
coupling. However, in all regimes the gate-voltage dependence of
the zero temperature conductance is found to be qualitatively the same
as in the model with no coupling to the vibrational mode.
\end{abstract}


\end{frontmatter}

In recent years the studies of quantum impurity systems have undergone
a considerable revival due to improvements in experimental techniques
for measuring the electron transport through quantum dots and single
molecules, as well as due to the development of the DMFT technique
which maps interacting lattice problems to quantum impurity problems
with an additional self-consistency condition. The prototype model for
this class of problems is the Anderson 
model for a single impurity in a metallic host with
$H_{\mathrm{imp}}=U n_{\uparrow}n_{\downarrow}  + \epsilon n$, where
$n=n_{\uparrow}+n_{\downarrow}$ is the number of electrons occupying
the impurity orbital with energy $\epsilon$
relative to the chemical potential, and with the Coulomb repulsion $U$
due to the double occupancy of the impurity orbital.

An important class of quantum impurity models include coupling to the
bosonic degrees of freedom describing the vibrational modes of the
molecule or phonons in the bulk. There are two basic types of the
electron-phonon coupling, (i) the Holstein coupling of form $n x$,
where $n$ is the electron density and $x$ the oscillator displacement,
and (ii) the coupling to hopping term of form $v x$, where $v$ is the
hopping (hybridization) operator that couples the impurity to the
conduction band. While type (i) is more relevant when the oscillation
is related to the change in volume to which the electron is confined
(breathing modes), type (ii) is relevant whenever the displacement
modulates the hopping probability.

The addition of the Holstein term to the Anderson Hamiltonian
effectively reduces the Coulomb repulsion and the hybridization $\Gamma$.
The effect of the electrons on the phonon propagator is also
interesting: when effective $U$ changes sign, a peak in the phonon propagator at
reduced frequencies (the 'soft mode') emerges
\cite{hewson02,jeon03}. The soft mode is related to the
charge susceptibility, which is increased in this regime  \cite{mravlje05}.

Very recently, similar behavior was found also in the case where the
electron-phonon coupling term is of the form $H_\mathrm{el-ph} = g x
v_{\mathrm{odd}}$, where $v_\mathrm{odd}$ describes the hopping from
impurity orbital to the odd conduction channel (antisymmetric
combination of the orbitals of the noninteracting part of the
Hamiltonian) \cite{balseiro06}.  The model without phonons consist of
$v_{\mathrm{even}}$, which couples the impurity only to the even
conduction channel (symmetric combination of orbitals of the
noninteracting part of Hamiltonian).

\begin{figure}
\begin{center}
\includegraphics[width=0.42\textwidth]{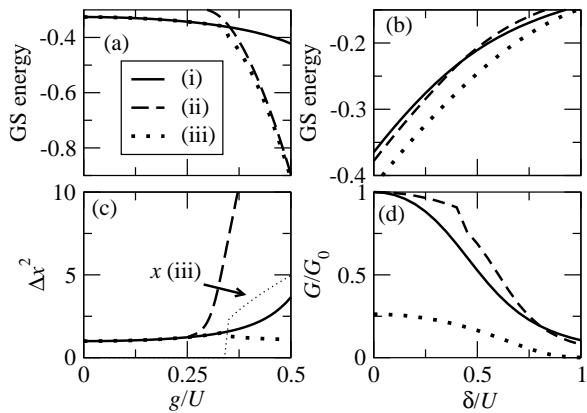}
\end{center}
\caption{Variational results [$U/\Gamma=5,
  \Gamma/D=0.04,\Omega=2.5\Gamma$; $\Omega$ is the phonon frequency, $D$
  the bandwidth. $\delta=0$ (left pannels), $g/U=0.36$ (right
  pannels)]: (a, b) Variational ground state energy. (c) Displacement
  fluctuations and displacement for (iii). In (i) and (ii) the parity symmetry is
  retained, hence the displacement vanishes. (d) The conductance.} \label{fig1}
\end{figure}

The same model (but for finite $U$ instead of $U\to \infty$ treatment
of Ref. \cite{balseiro06}) was analyzed also with the variational
projection-operator method \cite{mravlje06}. In this method the
ground state is expressed in terms of the ground state of an auxiliary
noninteracting Hamiltonian 
\cite{rr03}. Several variants were tested, with parameters chosen so
as to allow for coupling, (i) only to even channel, (ii) only to odd
channel, (iii) a combination of both. Variational method applied to
variant (iii) leads in certain parameter regimes to a ground state of
broken parity symmetry (see Fig. \ref{fig1}(a,b)), marked by
non-vanishing expectation value of displacement (Fig. \ref{fig1}(c), thin-dotted)
and consequently \cite{mravlje06} considerably reduced conductance
through a molecule ( plotted as a function of departure from particle-hole
symmetric point $\delta=\epsilon+U/2$ in Fig. \ref{fig1}(d)). As discussed below, the
ground state should have a well defined parity, therefore only the variants
(i) and (ii) of the variational procedure correspond to the ground
state of correct symmetry. While it would be instructive to implement the variational method in a
manner which correctly took into account the tunneling between the
classically degenerate minima of the oscillator potential
\cite{balseiro06}, it currently appears that the implementation would
require calculating matrix elements between two distinct Hartree-Fock
vacua, which precludes the use of Wick's theorem upon which our
current implementation of the variational procedure is based
\cite{rr03}.

\begin{figure}
\begin{center}
\includegraphics[width=0.4\textwidth]{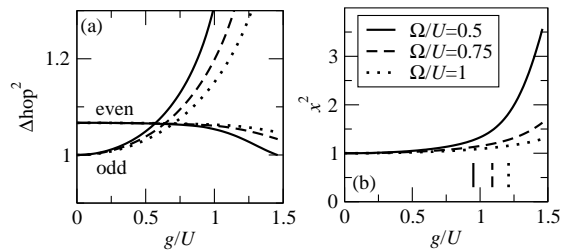}
\end{center}
\caption{NRG results [$U/\Gamma=25,D=U$]: (a) fluctuations of hopping to even (full) and odd (dashed)
  channel, (b) fluctuations of displacement.} \label{fig2}
\end{figure}

We thus present preliminary results obtained with the NRG method,
which does not suffer from this problem. The parity in NRG results is
not broken; the expectation values of the displacement
and of the hopping term $v_{\mathrm{odd}}$ thus vanish. In
Fig. \ref{fig2}(a) we compare the fluctuations of hopping to even and odd
channels as a measure of the 'activity' of corresponding channels.
For $g$ large enough the latter are larger, corresponding to increased
fluctuations of the displacement and the emergence of the soft mode.
The ground state of the system, as seen from the NRG renormalization
flow (not shown here) corresponds to the Fermi liquid ground state of
the single-channel Kondo problem \cite{hewson_book} with the
characteristic quasi-particle scattering phase shift
$\delta_{\mathrm{q.p}} \sim \pi/2$ in the
even or odd channel, depending on whether the effective phonon
mediated coupling to the odd channel is smaller or larger than the
direct coupling to the even channel. When the couplings match (marked
in Fig. \ref{fig2}(b) by vertical lines), an
unstable fixed-point of the two-channel Kondo model type is found.

\begin{figure}
\begin{center}
\includegraphics[width=0.28\textwidth]{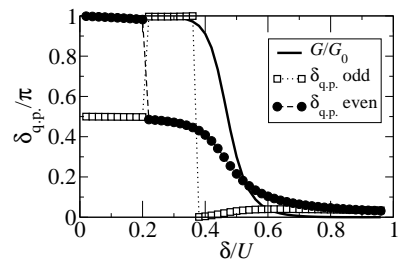}
\end{center}
\caption{NRG results [$U/\Gamma=25,D=U,\Omega/U=1,g/\Omega=1.3$]: Zero temperature quasiparticle scattering phase shifts in
  even (circles) and odd (squares) channel plotted versus departure
  from p-h symmetry calculated using NRG. Corresponding conductance is
  also plotted (dashed). Note that the phase shifts are defined modulo
  $\pi$.}\label{fig3}
\end{figure}

In Fig. \ref{fig3} the scattering phase-shifts are plotted as a
function of $\delta$. 
The coupling to the phonon-assisted channel is chosen so that in the ground
state the impurity spin is screened by the odd channel  for small $\delta$, while for larger  
$\delta$ it
is screened by the even channel, as seen from
$\delta_{\mathrm{q.p.;even,odd}} \sim \pi/2$, respectively. Further away from the 
symmetric point
the model is tuned into the valence fluctuating
regime where the Kondo effect is suppressed. When the model is used to
describe a molecule (or a quantum dot) embedded between two leads, the
scattering phase shifts directly determine the differential
conductance (Fig. \ref{fig3} dashed). The conductance curve shown is
qualitatively equal to that of the generic one-electron transistor in
the single-channel Kondo regime, despite the fact that the Kondo
effects occurs in different channels as the gate voltage is swept.
 
The work was supported by SRA under grant Pl-0044.

\end{document}